\documentclass[prl,twocolumn]{revtex4}

\begin{document}

\textbf{Mostovoy Reply:} In their Comment \cite{KenzelmannHarris}
Kenzelmann and Harris argue against the conclusion made in
\cite{Mostovoy2006} that spiral magnets are in general
ferroelectric. First of all, I believe, this conclusion was proved
experimentally. The systematic search for ferroelectricity in
magnets with spiral ordering recently led to a discovery of new
multiferroic materials, such as CoCr$_{2}$O$_{4}$ \cite{Yamasaki},
MnWO$_{4}$ \cite{Taniguchi,Heyer} and LiCu$_{2}$O$_{2}$
\cite{Cheong}.

Furthermore, Kenzelmann and Harris argue that the continuum theory
outlined in \cite{Mostovoy2006} leads to \emph{misleading
predictions} about the magnetically-induced electric polarization.
To prove their point, they consider two hypothetical spin
configurations shown in Fig. 1 (c) and (d) of their Comment, and
argue that the results of the continuum theory are incompatible
with crystal symmetries. While one cannot deny the importance of
symmetry considerations, the arguments Kenzelmann and Harris are
themselves very misleading. They incorrectly assert that for the
spin configurations shown in Fig. 1 (c) and (d) `the spiral
theory' would predict electric polarization along, respectively,
the $c$ and $a$ axes.

The continuum model of multiferroics \cite{Mostovoy2006} is based
on assumption that the spin state can be described by a single
magnetization vector. For TbMnO$_{3}$ (see Fig. 1b), where the
wave vector of the magnetic spiral is along the $b$ axis and spins
are rotating in the $bc$ plane, it predicts electric polarization
$\mathbf{P}$ along the $c$ axis, in agreement with experiment. The
magnetic structures (c) and (d) are of a different kind, as they
are made of spirals rotating in opposite directions. Thus in the
configuration (c) there are two counter-rotating $bc$ spirals in
each $ab$ plane, which is why the net polarization along the $c$
axis is zero. Similarly, in the configuration (d) the $ab$ spirals
in neighboring $bc$ planes rotate in opposite directions,
resulting in zero net $P_{a}$.

It is not difficult to modify the continuum model considered in
\cite{Mostovoy2006} to describe these more general magnetic
orders. For more than one magnetic ion per unit cell one can
introduce several independent magnetic order parameters, which
increases the number of possible magnetoelectric coupling terms.
For instance, all three spin configurations shown in Fig. 1 of the
Comment can be described by three antiferromagnetic order
parameters
\begin{equation}
\begin{array}{ccl}
\mathbf{L}_{1} & = &
\mathbf{S}_{1}+\mathbf{S}_{2}-\mathbf{S}_{3}-\mathbf{S}_{4}, \\
\mathbf{L}_{2} & = &
\mathbf{S}_{1}-\mathbf{S}_{2}+\mathbf{S}_{3}-\mathbf{S}_{4}, \\
\mathbf{L}_{3} & = &
\mathbf{S}_{1}-\mathbf{S}_{2}-\mathbf{S}_{3}+\mathbf{S}_{4}
\end{array}
\end{equation}
(the labels of the 4 Mn ions in the unit cell of TbMnO$_{3}$ are
the same as in \cite{HarrisLawes}). The spiral configuration (b)
can be described by a single order parameter $\mathbf{L}_{1}$ with
nonzero $L_{1}^{b}$ and $L_{1}^{c}$. As discussed in
\cite{Mostovoy2006}, the magnetoelectric coupling linear in the
gradient of the magnetic order parameter (Lifshitz invariant)
allowed by symmetries has the form $P^{c}\left( L_{1}^{c}\partial
_{y}L_{1}^{b}-L_{1}^{b}\partial _{y}L_{1}^{c}\right) $, which
gives rise to magnetically-induced $P^{c}$. The configuration (c)
is described by two different order parameters, $L_{1}^{b}$ and
$L_{3}^{c}$. The term $L_{3}^{c}\partial
_{y}L_{1}^{b}-L_{1}^{b}\partial _{y}L_{3}^{c}$ does not transform
like any of the components of $\mathbf{P}$, so that the induced
polarization is zero. Finally, for the configuration (d) with
nonzero $L_{1}^{b}$ and $L_{2}^{a}$, the only possible coupling
term is $P^{c}\left( L_{1}^{b}\partial
_{y}L_{2}^{a}-L_{2}^{a}\partial _{y}L_{1}^{b}\right)$, allowing
for nonzero $P^{c}$.

The point is, however, that the spin configurations (c) and (d)
considered by Kenzelmann and Harris, are very artificial, as it
is difficult to find a system where interactions between spins
would favor the simultaneous presence of counter-rotating
spirals. The average interaction between counter-rotating
spirals is zero, while for spirals with spins rotating in the
same direction some interaction energy can always be gained by
properly adjusting their relative phases. This is the reason why
the simple model of Ref. \cite{Mostovoy2006} with a single
vector order parameter successfully describes thermodynamics and
magnetoelectric properties of many spiral multiferroics.

\flushleft{Maxim Mostovoy\\ Materials Science Center, University
of Groningen, Nijenborgh 4, 9747 AG Groningen, The Netherlands}

\end{document}